# Enriched multi-agent middleware for building rule-based distributed security solutions for IoT environments

Francisco José Aguayo-Canela[1] · Héctor Alaiz-Moretón[1] ·
María Teresa García-Ordás[1] · José Alberto Benítez-Andrades[2] ·
Carmen Benavides[2] · Isaías García-Rodríguez[1]

## Abstract
The increasing number of connected devices and the complexity of Internet of Things (IoT) ecosystems are demanding new architectures for managing and securing these networked environments. Intrusion Detection Systems (IDS) are security solutions that help to detect and mitigate the threats that IoT systems face, but there is a need for new IDS strategies and architectures. This paper describes a development environment that allows the programming and debugging of distributed, rule-based multi-agent IDS solutions. The proposed solution consists in the integration of a rule engine into the agent, the use of a specialized, wrapping agent class with a graphical user interface for programming and debugging purposes, and a mechanism for the incremental composition of behaviors. A comparative study and an example IDS are used to test and show the suitability and validity of the approach. The JADE multi-agent middleware has been used for the practical implementations.

## 1 Introduction

The proliferation of devices with Internet connection capabilities in the so-called Internet of Things (IoT) is a trend that is generating an overwhelming amount of new streams of data. These data are crucial to the operation of the systems where the devices are located but must be properly managed to obtain useful information for decision-making. The distributed nature of these systems demands decentralized architectures for the management and control of the IoT devices, including tasks

✉ José Alberto Benítez-Andrades
  jbena@unileon.es

Extended author information available on the last page of the article such as monitoring or security assurance [1]. The multi-agent paradigm has proved to be a convenient approach for building this kind of decentralized management and control systems [2]. Regarding cybersecurity issues, the complex nature of IoT ecosystems demands integrated defending systems that can cope with different threats which such systems may be exposed to.

One of the needs for IoT ecosystems protection is to build tools able to detect malware activity within a given network as soon as possible, in order to minimize the number of devices affected or the extent of the damage that could be produced. Intrusion Detection Systems (and Network-based Intrusion Detection Systems (NIDS) in particular) are a family of technological solutions that have traditionally being used with this purpose. In IoT scenarios, the number of different devices, protocols and communication acts take these security challenges to a new level. The use of intelligent, cooperative agents has proved to be a convenient solution for defending such IoT ecosystems [3]; different tasks involved can be distributed among a number of specialized software entities that cooperate by exchanging messages in order to obtain a common goal. This approach has led to a shift in (N)IDS from monolithic and very centralized solutions to distributed ones, who are known as Collaborative Intrusion Detection Systems (CIDS) [4].

The "intelligence" provided by the agents in order to build their intrusion detection capabilities may be based on the use of rules, different sorts of logic and other deliberative mechanisms and AI techniques [5]. Many existing cognitive functions used for building intelligence into the agent are based on the reactive model, using an event-driven mechanism that, with the aid of a set of rules, allows the agent to trigger its behavior when a given condition is detected. This mechanism is specially useful for building the so-called misuse or signature-based intrusion detection systems, as they use the known characteristics of malicious traffic in the network to recognize potential threats. The usual formal representation for this kind of knowledge consists in a set of rules that are fired when their activation patterns match the monitored traffic features, but building such rule-based reactive systems into the agents of a multi-agent system (MAS) is a difficult, time-consuming task.

This research aims at designing and building a development environment for easing the construction and debugging of multi-agent-based security systems that use rules for implementing the cognitive capacities of the agents. The solution is based on a set of tools and functionalities incorporated into the multi-agent middleware that eases the design, implementation, test and development of this kind of distributed NIDS systems.

The solution proposed includes a loosely coupled integration of the rule engine into the agent, not depending on the implementation technology. The development and deployment of the agents in the multi-agent platform will be incremental, with aids for testing not only the own agent's knowledge base, but also the cooperative problem solving involving different agents.

The multi-agent middleware chosen for testing this proposal is JADE [6]. This choice is based on its maturity, the size of the user community and the use of behavior-based agents, which can be exploited to integrate the rule engine processes as behaviors.

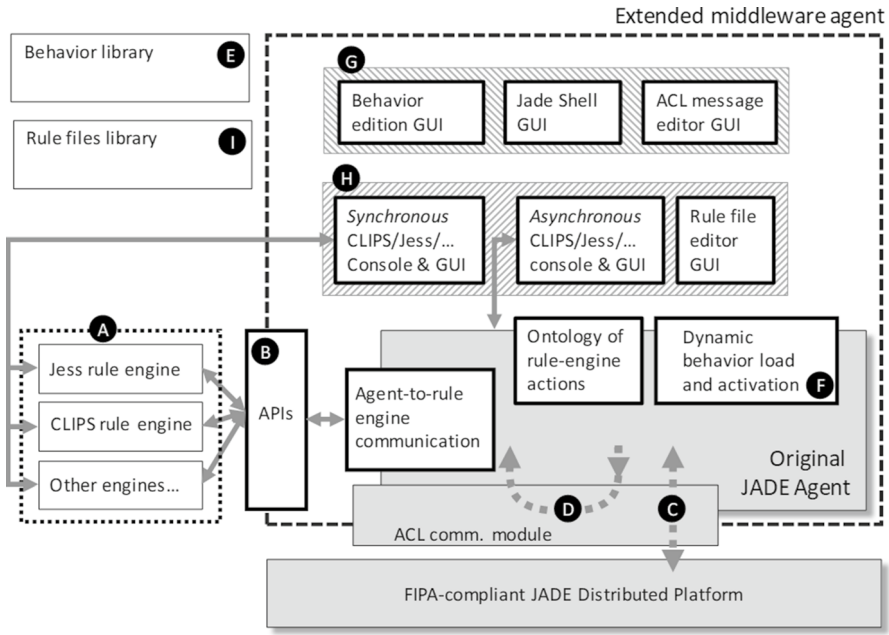

**Fig. 1** Framework for the proposed solution

Figure 1 shows a diagram of the proposed middleware architecture. The rule engine (A) used by a given JADE agent is loosely coupled to it by means of a software interface (B) according to the technology of the engine. The rule-based agent will communicate with other rule-based agents by using the standard Agent Communication Language (ACL) (C) and a set of technology-neutral concepts for describing actions in the rule engines that is stored in a shared ontology (see Table 1). As well as the agent-to-agent, also the agent-to-rule engine communication is achieved by using ACL messages that the agent sends to itself (D).

Besides the integration of the agent and the rule engine, the proposed solution includes a set of facilities in order to ease the design, development and debugging of rule-based multi-agent systems. The basic JADE Agent class has been extended in order to build an incremental behavior composition system by using a library of externalized behaviors (E) and a mechanism for the dynamic load and incorporation of these behaviors into the agent (F). This functionality is detailed in Sect. 4.

A number of graphical user interface windows have also been built into the extended JADE Agent class to aid the programmer in the process of building the knowledge base for each agent and test the distributed multi-agent system prior to put it into production. The "agent management tab" (G) includes a behavior edition window, a JADE Shell editor and an ACL message visualization and edition window; all of them are described in Sect. 3.2. On the other hand, the "rule engine management tab" (H) includes a file editor for creating and editing rules that are stored in an external library (I), a synchronous shell for direct interaction with the given rule engine used by the agent and an asynchronous shell where the user can interact both with the rule engine of the same agent and with any rule engine of any agent in the platform. Section 3.3 gives further details about these windows and their functionalities.

Table 1 Codes and description for possible activities to be communicated between rule-based agents

| Code | Description |
| --- | --- |
| LOAD_FILE | Load the file indicated as a parameter |
| LOAD_FACTS | Loads the facts file indicated as a parameter. |
| LOAD_FROM_RESOURCE | Loads the given resource file indicated as a parameter. |
| LOAD_FROM_STRING | Loads data from a CSV file indicated as a parameter. |
| LOAD_ASSERT_STRING | Loads facts from a string. |
| LOAD_BLOAD | Memory restoring from a bin file. |
| LOAD_SLOAD | Memory restoring from a plain text file. |
| RUN_INFINITELY | Run indefinitely up to the end of rule activations. |
| RUN_NUMBER_OF_CYCLES | Run a given number of cycles. |
| RUN_ONCE_THEN_BATCH | Run and give the control back to the Shell. |
| RUN_INNER_SHELL | Execute the internal Shell. |
| MAKE_RESET | Perform a reset command. |
| MAKE_CLEAR | Perform a clear command. |
| MAKE_MEMORY_DUMP | Perform a security backup. |
| MAKE_ASSERT_STRING | Inserts a fact from a string. |
| MAKE_BUILD | Compile a query. |
| EVAL_COMMAND | Evaluate a sentence. |
| SET_INPUT_BUFFER_COUNT | Requests the number of input characters entered. |
| APPEND_INPUT_BUFFER | Appends to the given command. |
| SET_UNWATCH | Not to analyze (debugger). |
| SET_WATCH | Analyze (Facts, Modules, etc.). |
| GET_FACT_SLOT | Get an slot value. |
| FACT_INDEX | Move the cursor in the fact list. |

The rest of the paper is organized as follows: Sect. 2 describes the integration of a rule engine into the agent, showing its main features and a test for validating and comparing the solution to another similar one. Sections 3 and 4 present the enriched multi-agent middleware and the externalization of the agent behaviors, respectively, while Sect. 5 shows how the proposed solution can be used as a development environment for building an example rule-based Network Intrusion Detection System. The objective of Sect. 5 is to show and demonstrate the features of the enriched middleware to ease the construction of such systems, not to build a fully functional NIDS solution. Some discussion is presented in Sect. 6 and, finally, Sect. 7 is devoted to show the conclusions and future work.

## 2 Integrating a rule engine into an agent: the rule-based agent

The integration of the rule engine into the agent is the base of the proposed enriched middleware; a detailed description of this solution can be found in [7]. The resulting integrated solution includes the following characteristics:

– Neutrality concerning the particular technology of the rule engine, for example, CLIPS [8], JESS [9] or Drools [10].
– The agent is the only responsible for the actions performed in its rule-based system, controlling the execution of the rule base independently from other actions, activities or behaviors. The rule engine is exclusively devoted to the agent, what distinguishes this approach from other ones where the reasoning is built as a service in a special agent devoted to solely perform the execution of the rules that other agents demand [11]
– The rule engine associated with an agent does not block the basal agent behavior while reasoning, allowing it to keep communicating with other agents.
– The design of the solution prioritizes the ease for the design and development of rule-based multi-agent applications.

The communication mechanism between agents with integrated rule engines complies with the Foundation for Intelligent Agents (FIPA) specifications, using ACL messages and a domain ontology for storing the valid set of message contents. The set of actions that a rule-based agent can ask another one to perform on its rule engine is limited to a predefined set of concepts (see Table 1) that represent the usual activities of these kind systems (loading facts and rules, executing a number of firing cycles, query facts and rules, performing a reset or clear command, etc.).

These actions can be invoked by a *rule-based agent* when communicating to another rule-based agent, but also by a human (working within the enriched interface shell of an agent, see Sect. 3.3.3) that wants to communicate with any other *rule-based agent* in the platform for development or debugging purposes. The size of the ACL messages exchanged by the agents ranges from 2 to 5 kBytes, being the bigger ones those who carry a set of fact or rule assertions in their payload.

### 2.1 Validation of the rule engine integration

In order to validate and test the proposed integration approach, a comparative study has been designed and implemented. The study compares the solution described in this paper to the integration described in [12]. The study is detailed and can be reproduced by following the indications and using the software at https://secomuci.com/research/MAS/IMAS/validation.

The study compares the performance and response times of an agent from the framework proposed in this paper (referred to as *DPSNodeAC*) and a *JessAgent* from the solution proposed in [12] (named *HLCjessAgent* in this section). The experiment consists in making a third agent, called *Analyzer*, generate a number of messages to be sent to the agents under test (*HLCjessAgent* and *DPSNodeAC*). The messages can be of two types:

– Presence request messages (used for testing if the agent is alive in the platform). The response is a simple acknowledgement for confirming the presence of the agent. The usual response time for this kind of message in the mentioned computer was about 300 ms when the agent is free from other reasoning processes.
– Requests messages asking for solving sudoku problems of different difficulties. The response to these messages is the solution found for the given problem and so they take the agent more time to respond than the presence request messages. There are four different sudoku problems to be solved, with solving times (in the rule base) from around 200 ms to 2500 ms.

The sequence of messages used is shown in table 3. A total of 40 messages were sequentially generated for each agent. The first four are of type "presence request" (p in Table 2), the fifth is of type "sudoku" (S in Table 2), then nine more "presence request" messages are sent and one "sudoku" follows; this sequence is repeated twice, ending with five more "presence" messages. Each message is scheduled to be sent from the *Analyzing* agent every 250 ms. So, the entire message sequence is generated within an interval of 10 seconds.

The *Analyzer* agent is responsible for sending the messages and capturing the corresponding responses, annotating the time at which the message was issued and the time when the corresponding response from the agent arrived; the difference is the corresponding delay for the given message.

Figure 2 shows the delays in the responses for each message for the *HLCjessAgent* (a) and for the *DPSNodeAC* agent (b). As can be seen, the approach described in this paper outperforms the results of the other solution. This improvement is achieved by the implemented integration of the rule engine into the agent that, in the case of the proposed solution, does not block the communication processes of the agent, as opposed to the compared solution.

**Table 2** Sequence of messages for the experimental test of the solution

| Message# | 1 | 2 | 3 | 4 | 5 | 6 | 7 | 8 | 9 | 10 | 11 | 12 | 13 | 14 | 15 | 16 | 17 | 18 | 19 | 20 |
|---|---|---|---|---|---|---|---|---|---|---|---|---|---|---|---|---|---|---|---|---|
| Type | p | p | p | p | S | p | p | p | p | p | p | p | p | p | S | p | p | p | p | p |
| Message# | 21 | 22 | 23 | 24 | 25 | 26 | 27 | 28 | 29 | 30 | 31 | 32 | 33 | 34 | 35 | 36 | 37 | 38 | 39 | 40 |
| Type | p | p | p | p | S | p | p | p | p | p | p | p | p | p | S | p | p | p | p | p |

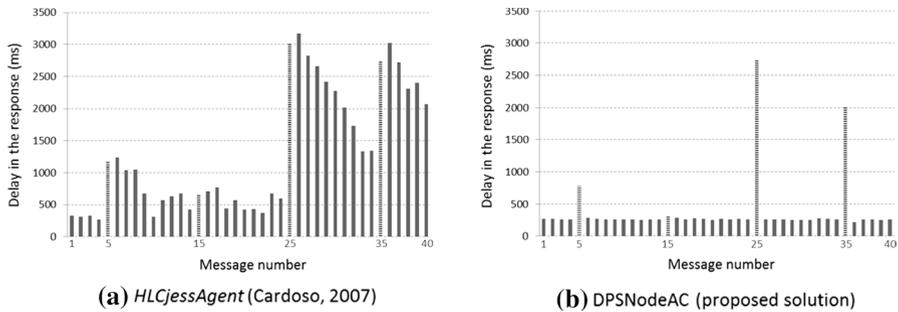

**Fig. 2** Delays in the responses of HLCjessAgent vs DPSNodeAC

## 3 Description of the enriched middleware

As well as the integration of the rule engine into the agent, the other main contribution of the proposed solution is the enriched multi-agent middleware to help in the development and test of rule-based multi-agent systems. The next sections are devoted to present different components of this enriched middleware.

### 3.1 The development environment

A core component of the solution is the "*development environment*," built as an agent extending the basic *Agent* JADE class. It is primarily intended to ease the development and debugging of agents with rule-based behaviors. When in development and debugging time, the *rule-based agent* must be invoked as a "*development environment*" agent in the platform. This invocation causes the agent to launch with an enriched graphical user interface with a set of functionalities for building and testing *rule-based agents*.

This solution allows the interactive modification of the agent internal code, as well as the interaction with its own rule engine, or even with any other rule engine in any of the agents in the platform. The graphical interface of the *development environment* agent has two main tabs (see Fig. 3). Tab 1 is called the "agent management tab" and tab 2 is the "rule engine management tab." Their functionalities are presented in next sections.

### 3.2 The agent management tab

The agent management tab includes the following windows:

- The JADE shell window.
- The behavior editing window.
- The message viewing and editing window.

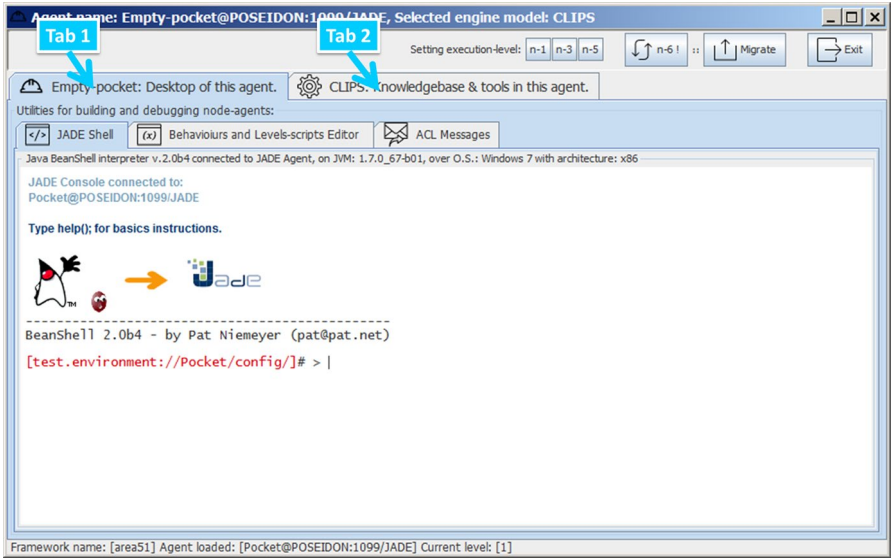

**Fig. 3** Graphical user interface of the development environment

### 3.2.1 The JADE shell window

The JADE shell window (see Fig. 4) contains the BeanShell component by [13] connected to the agent to allow a human to program in Java and having direct access to the JADE API objects, the methods and properties of the agent, the methods and properties of the rule engine, etc. It allows to build new classes and to instantiate new objects to be incorporated into the agent, to look at the help files, to perform unitary tests, to watch the message queue, to build new messages and send them, to build and test new behaviors, etc.

The BeanShell component also has a non-graphical mode that is used when the agent is in production (without this graphical user interface) to read and load the behaviors of the agent at runtime (see Sect. 4).

### 3.2.2 The behavior editing window

The behavior editing (see Fig. 5) contains a text editor making use of the RSyntaxTextArea component (https://bobbylight.github.io/RSyntaxTextArea/). It is used for loading and modifying the behavior files of the agent (see Sect. 4). It has syntax highlighting and word auto-completion capabilities. On the left part of the window, a list shows the set of all the possible behavior files for the agent in order to be loaded into the editor.

**Fig. 4** JADE shell window

**Fig. 5** Behavior editing window

### 3.2.3 Message editor and trace window

The message editor window (see Fig. 6) contains a partial implementation of the *testAgent* component in the *jade.tools.testagent* package distributed with the JADE middleware. It allows to watch the events in the message queue of the agent and manually build new messages.

## 3.3 The rule engine management tab

The "rule engine management tab" includes the functionalities to interact with the associated rule engines of the multi-agent system. It is composed of three windows:

– The file editor, for the expert system managed by the agent.
– The synchronous shell, for communicating with the local rule engine.
– The asynchronous shell, for communicating with any remote rule engine in another agent in the platform.

### 3.3.1 File editor window

Figure 7 shows the file editor window. It is used for editing expert system files locally. It is based on the *RSyntaxTextArea* component and includes CLIPS and Jess syntax highlighting. The files created here can be later loaded into the agent's rule engine or sent remotely to another *rule-based agent*.

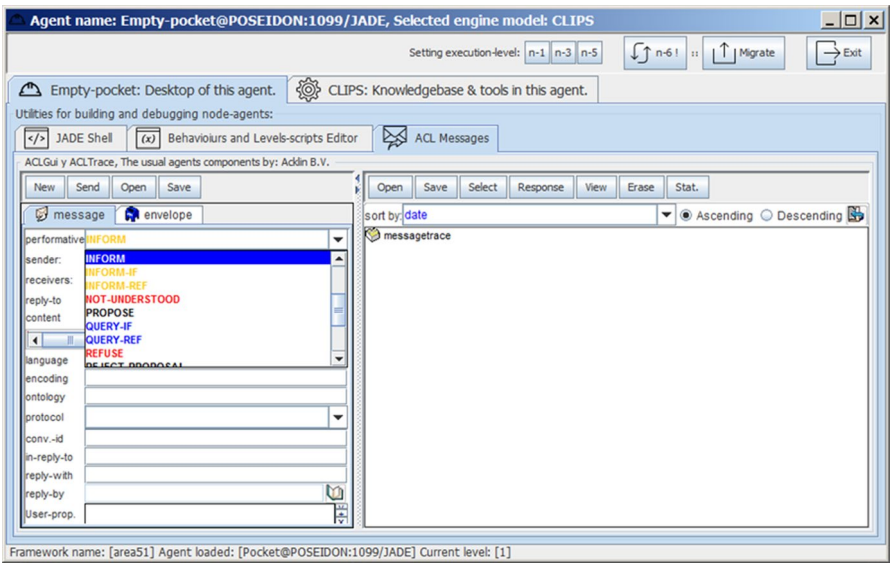

**Fig. 6** Message editing window

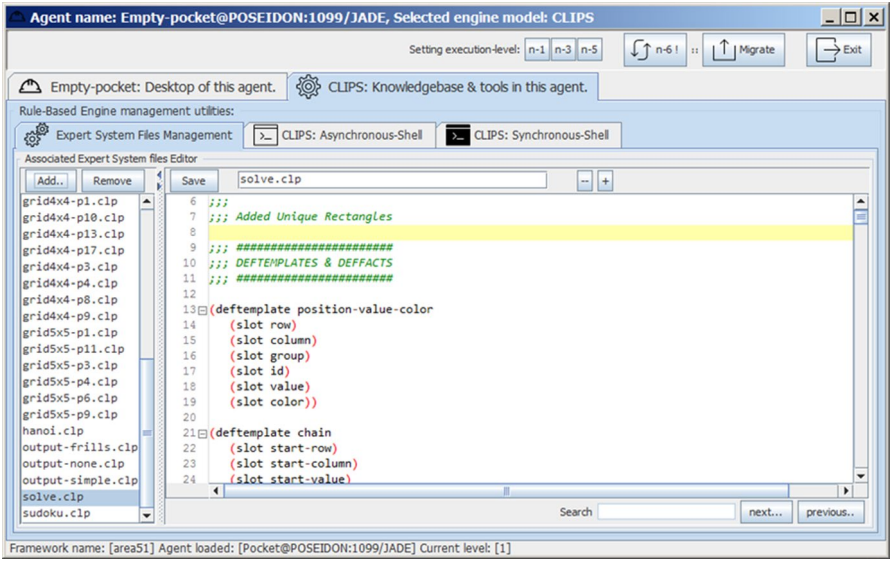

**Fig. 7** File editing window

### 3.3.2 The synchronous shell window

The synchronous shell allows a human to interact directly with the rule engine of the agent. It performs a direct connection from the graphical interface to the inner rule engine, emulating a shell of the underlying technology (CLIPS in the case of Fig. 8). This window should only be used during the initial phases of the agent development, and not during execution, where the asynchronous shell is preferred for not to block the operation of the agent.

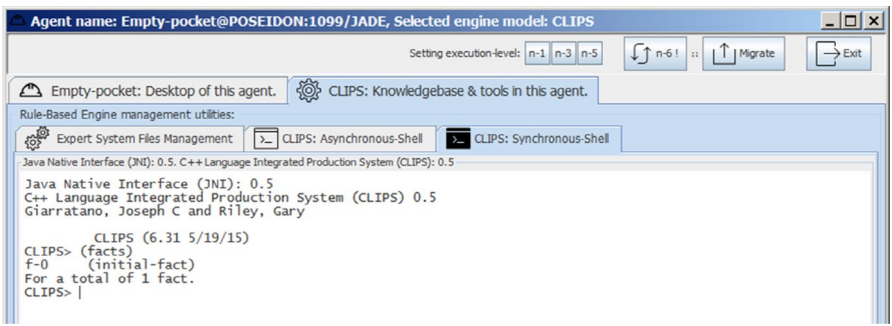

**Fig. 8** Synchronous shell window

### 3.3.3 The asynchronous shell window

This shell allows the communication with the rule engine of any agent in the platform, including the one of the agent where the interaction takes place (see Fig. 9). The command introduced in this window is included in an ACL message that is sent to the rule engine of the destination agent. It allows the communication with the agents and their rule engines at runtime, using an interaction protocol, without blocking the agent behaviors or the engine execution.

Once the command is introduced in the upper text area, it can be dispatched by using the combination Shift+Ctrl+Enter, or clicking on the "Execute!" button. The list in the left allows selecting which of the agents in the platform will be the destination agent, including the own local agent (denoted by the word "itself" in the list). The command is sent to the destination agent in an ACL message and, once the rule engine of that agent processes the instructions, the results are sent back to the sender in an ACL *Notification* message associated with the conversation thread created at the beginning of the interaction. There is a text area at the bottom of the window where the responses of the destination agent (usually the results of the processing of the instructions by the rule engine) will appear. As a result of the solution designed, the window is not blocked while waiting for the response, neither is the destination agent. So, new commands can be sent even to the same destination agent.

## 4 Externalization and incremental composition of behaviors

The externalization of behaviors allows an agent to load or modify its behaviors by loading and processing them in real time from local files. The Java interpreter, incorporated into the agents (see Sect. 3.2), is responsible for the processing of these external files and the incorporation of the behaviors in the task manager at a proper time to avoid collisions and incoherences.

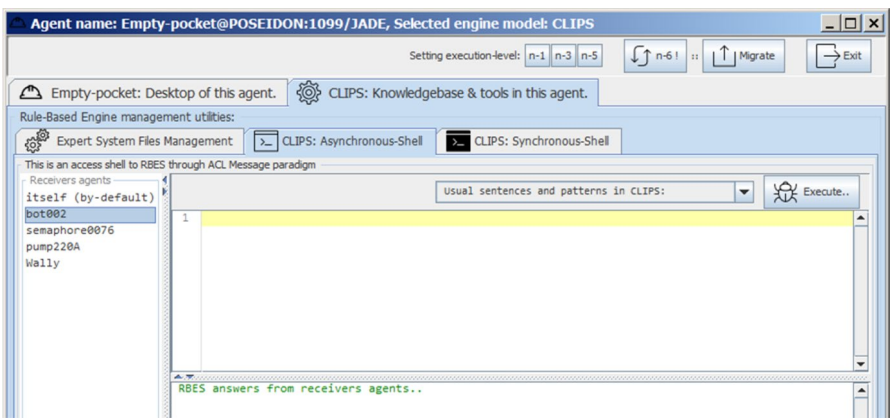

**Fig. 9** Asynchronous shell window

Table 3 Runlevels and corresponding actions

| Level | Process |
|---|---|
| 0 | The setup() method for the agent finished its execution. The agent is already incorporated into the multi-agent platform and its status is active. The script file [level.00.bsh] is loaded and interpreted. |
| 1 | The script file [level.01.bsh] is loaded and interpreted, which results in the load of the "basal" behaviors for the agent. |
| 3 | Load and interpretation of the script [level.03.bsh]. Activation of behaviors loaded in level [1], objects of the type Behavior that appear in the behavior collection are also loaded. |
| 5 | Load and interpretation of the script [level.05.bsh]. Activation of the behaviors that were loaded in level [3]. Whenever the scripts in [level.05.bsh] are processed, the agent is considered in the state "in service," and the execution level is set to [5]. |
| 6 | The script [level.06.bsh] contains commands that result in a "hot reboot" of the agent, that means that the agent is not removed from the platform, but its active behaviors are stopped and removed from the agent. Following, the execution level [0] is entered. |

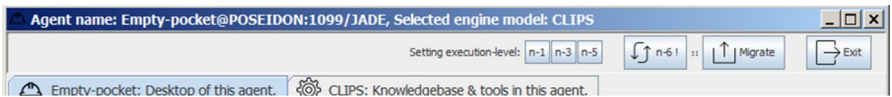

Fig. 10 Buttons for runlevel activation

This dynamic process of loading the agent behaviors is the base for the incremental composition of behaviors mechanism. The final, complete, behavior of the agent can be composed of different behaviors that can be loaded one at a time. This way, the behavior can be tested step by step, starting with the simplest or basal ones. To ease this modularity and progressivity in the construction of the final behavior, the agents are initialized by going through a series of steps very similar to the "runlevels" found in UNIX-like operating systems.

Five runlevels are defined; each of them has an associated script file associated where the behaviors to be loaded are indicated. Table 3 shows different runlevels and the associated processes that occur in each of them.

The execution level of the agent can be controlled with the buttons "n-1," "n-3," "n-5" and "n-6!" in the graphical user interface (see Fig. 10).

In practice, these runlevels can be used to incrementally test the functionalities of the agents, for example when building complex behaviors or when testing coordination mechanism with other agents.

# 5 An application example scenario: distributed intrusion detection system

As stated in the introduction, the use of multi-agent systems for building intrusion detection systems has a number of advantages; the distribution obtained by using MAS technology helps to enhance the detection accuracy and makes a more effective system [4].

This section is devoted to show how the proposed enriched multi-agent middleware helps to develop a simple example NIDS that could be used for detecting malicious traffic in the network. As stated earlier, this presentation has the objective of showing different features built in the enriched multiagent middleware and how can they ease the development of these security solutions. First, different types of multi-agent-based NIDS will be briefly presented. Then, two of them, using a rule-based formalism, will be described to serve as a guide for designing the proposed example NIDS. Finally, this example NIDS will be presented and the its agents will be described, showing how the enriched middleware helps in their design, development and debugging.

## 5.1 Different types of multi-agent NIDS

There are a number of different approaches to build multi-agent-based (N)IDS; they differ in the agent architecture as well as in the techniques and algorithms used to perform the agent reasoning [4]. Some of the solutions are implemented by knowledge-based systems using rules to implement the agent cognitive functions [14, 15]; other approaches use machine learning techniques [16], artificial neural networks [17] or (more recently) deep learning techniques [18]. It is also usual to find hybrid systems using more than one of these approaches.

The implementation to be described next would be included in the category of knowledge-based NIDS, where the detection capabilities of the agents are based on the use of known features of the monitored network traffic considered as malicious. These features are modeled in an explicit or implicit conceptual structure, while the reasoning is performed by means of rules that use patterns resembling the traffic feature values the IDS is looking for.

The solution presented in [14] describes a multi-agent system with the following agents:

– Monitor Agent is responsible for monitoring activities and host statuses. It captures network packets and forwards them to the analysis agent for further processing.
– Analysis Agent, this agent has multiple instances running at the same time. It processes the traffic with the aid of a knowledge base, deciding if it is malicious or not. If it is malicious, then the Executive Agent is called.
– Executive Agent, it is in charge of warning other nodes in the distributed IDS. Also when the Analysis Agent detects malicious packets, but the pattern of intru-sion is new, then the Executive Agent adds this new pattern to the Knowledge Base.
– Manager Agent. The function of this Manager Agent is to supply information of host performing malicious activities to the Analysis Agent.

The agents use a common knowledge base component, consisting of attack patterns, malicious code patterns that are used by the analysis agents and updated by the executive agent. That means it contains certain rules which help the Executive Agent decide upon as to which data are malicious and thus generate warning. Authors do not provide any information about the practical implementation of the approach or the technologies used.

The OMAIDS system [15] uses an ontology for explicitly modeling the knowledge of the domain. The systems detect the attacks through the intelligent MisuseAgent agent, which uses the ontology to enrich data intrusions and attack signatures by semantic relationships. The multi-agent system is comprised of the following agents:

– The SnifferAgent captures packets from the network, preprocesses them and send the results to the MisuseAgent.
– The MisuseAgent receives the packets from the Sniffer-Agent. It transforms these packets to OWL (Ontology Web Language) format in order to be compatible with the Semantic Web Rule Language (SWRL) rules stored in the ontology. If there is a similarity between the OWL packets and the SWRL rules that define the attack's signatures, then the agent raises an alert to the ReporterAgent.
– The ReporterAgent generates reports and logs.

Authors used the JADE multi-agent platform for their implementation, but do not provide information about the solution used for integration of the rule engine into the agent or other technological details.

### 5.2 The example NIDS solution

The proposed NIDS solution to be used as a demonstrator for the enriched multi-agent middleware uses three different types of agents:

– Watchdog agents are responsible for processing previously captured network traffic pcap files and converting them to fact files to be used by expert system agents.
– A NIDSBoardAgent is responsible for loading fact files into a shared knowledge base, maintaining and announcing the state of the capture files (whether if the capture is being, or has been, processed by any agent in the platform) and rising the corresponding notification when an alarm is raised by the rule-based expert system agents
– Rule-based expert system agents are responsible for analyzing the capture files, comparing the capture features against different rule patterns they have, and sending messages to the NIDSBoardAgent for updating the alarm status (indicating if an alarm must be notified, for example).

The enriched middleware proposed in this research allows the incremental implementation of each of the mentioned agents and their interactions. Moreover, it allows the debug of the agents in real time when they are running.

The following subsections show different functionalities of the enriched middleware in the context of the construction of the previously described NIDS agents. The code and detailed instructions for reproducing this experiment can be downloaded and consulted at https://secomuci.com/research/MAS/IMAS/NIDS

### 5.2.1 The NIDSBoardAgent

The NIDSBoardAgent is a specialized agent responsible for maintaining the knowledge base of network traffic captures, along with the state information about the processing status and the results of the analysis notified by the processing agents (rule-based expert system agents). The enriched middleware for this type of agent includes three tabs when being used in development time (see Fig. 11 -split into two for readability purposes-): the *Ticket's Repository Table*, the *Behaviors and Execution-level Editor*, and the *JADE Shell*. The *repository table* tab includes a view of the knowledge base containing the information for each of the network traffic captures:

– The TIDREPLY column contains the ID of the conversation between this agent and the corresponding rule-based expert system agent that requested the capture for analysis.
– The DEWEYCODE column contains identification for the type of packets that the capture comprises; these data can be used for specializing the rule-based expert system agents, allowing them to retrieve only the data they can analyze.
– The STATEUNTIL column contains the analysis status as notified or stated by the rule-based expert system agent:
  – checkout: the capture is being processed by the corresponding rule-based expert system.
  – aborted: the analysis finished with an error.
  – ALERT !: the analysis finished and contains one or more alerts.
  – finished: the analysis finished with no malware patterns detected.
– Other columns contain, among others, the name of the agent processing the corresponding capture (PARMETHOD), the firing strategy used by the corresponding rule engine of the agent that performed the analysis (KEYMETHOD) or the name of the file containing the rules used by the corresponding agent for the analysis (ENGINE). All this information is very useful for testing and debugging purposes as well as fine-tuning the NIDS operation.

The *JADE Scripting Shell* tab allows the access to all the internals of the NIDSBoardAgent. The agent variables can be obtained and manipulated using

**Fig. 11** Ticket's Repository Table of NIDSBoardAgent

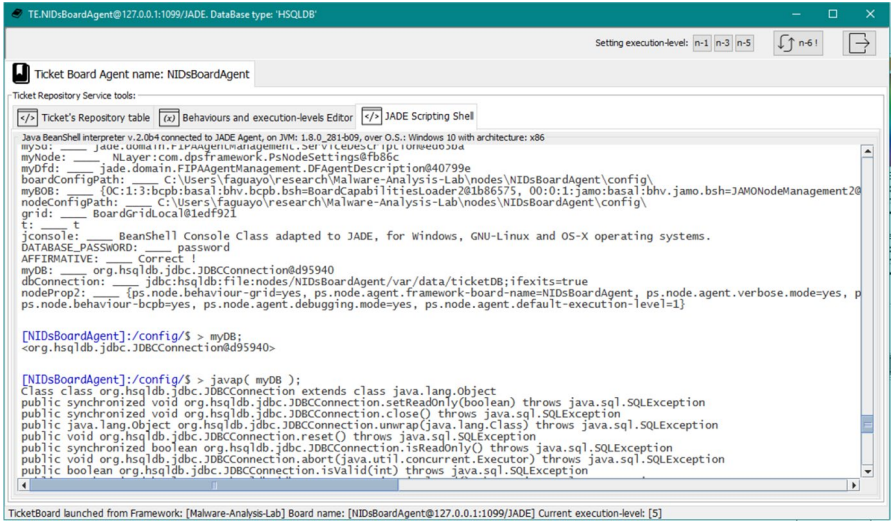

**Fig. 12** JADE Scripting of ShellNIDSBoardAgent

shell commands, including the database containing the knowledge base of the NIDS (see Fig. 12).

The *Behaviors and execution-level Editor* tab (Fig. 13) allows the incremental programming of the NIDSBoardAgent agent, and the modification of its behaviors in run time.

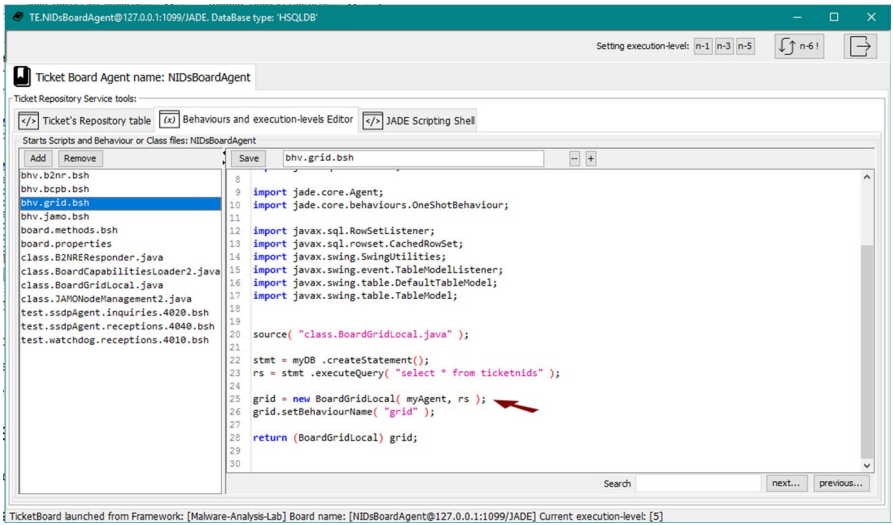

**Fig. 13** Behaviors and execution-level Editor of ShellNIDSBoardAgent

Different functionalities can be associated with different run levels and so the agent can be put into different working states in order to incrementally test its functionalities. Once the agent is ready, it will be invoked in the MAS platform without this enriched set of tabs.

### 5.2.2 The rule-based expert system agent

The rule-based expert system agents have, during developing time, three tabs devoted to the management of its integrated expert systems and two more tabs for managing its behaviors and properties. Figure 14 shows the tab devoted to editing different files that will form the expert system of the agent. By clicking on the corresponding file on the left, the rules can be edited and managed. The CLIPS shell tabs can be used to test the rule-based expert system of the agent by simulating the existence of given set of capture packets inserting them as facts in the agent knowledge base. The agent rules can be tested at run time by putting it into an execution level that does not include the communication capabilities with other agents in the platform, thus isolating it from the rest of agents until the expert system is ready to be put into production.

Different behaviors for each of the run levels of the agent can be programmed and tested using the behavior editing tab for the agent (see Fig. 15). The agent has a JADE shell in order to access and edit its internal variables and source code. When the behavior of the agent is modified it can be restarted by a direct click on the n-6 button or by incrementally setting the run level from n-1 to n-6 with the corresponding buttons.

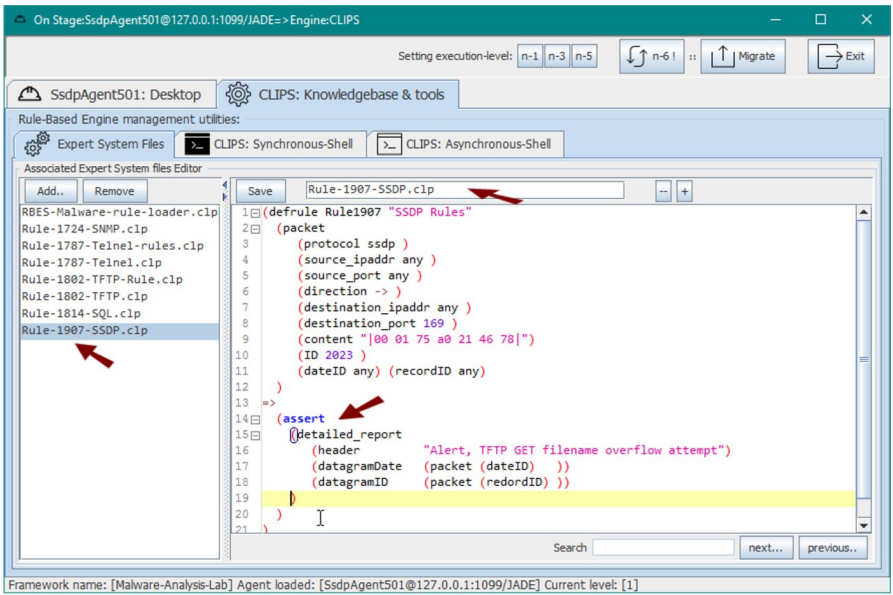

**Fig. 14** File editor window of a rule-based expert system agent

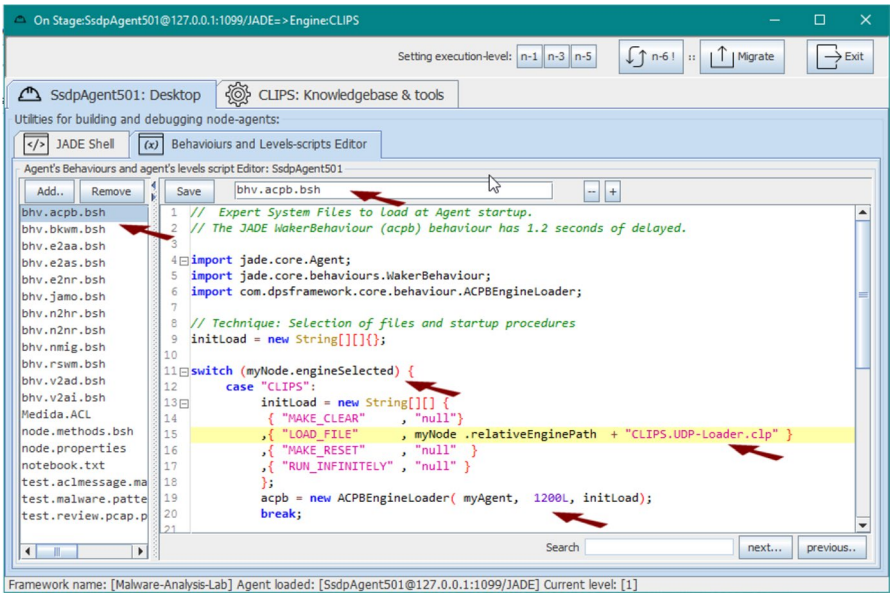

**Fig. 15** JADE shell window of a rule-based expert system agent

## 6 Discussion

This work describes an enriched middleware solution for multi-agent platforms with the aim of easing the development of distributed, rule-based, security solutions for Internet of Things scenarios. The solution is based on an integration of the rule engine into the agent and an enriched development environment that is built by extending the MAS platform agent components with a series of extra functionalities and graphical interfaces.

The integration of the rule engine into the agent is achieved by means of a loosely coupled strategy, implementing a mechanism for the communication of the agent with its rule engine that allows the agent not to be blocked while performing reasoning in its rule engine and so being able to keep receiving messages from other agents in the platform. This integration outperforms the usual ones found in similar approaches (as shown in Sect. 2) that, while performing the reasoning in the rule engine, block the functions of the agent preventing it from keeping communicating with other agents in the platform. The integration proposed in this research is crucial for improving the working speed of agents when they are part of an Intrusion Detection System.

By means of the example NIDS that has been built, different features of the enriched middleware have been presented. These features convert the multi-agent platform and the agent component into a complete development environment where agents can be designed, incrementally programmed, deployed and tested.

This research is centered in building rule-based multi-agent systems. Today, the trend about NIDS is to use machine learning and deep learning algorithms and techniques for implementing the agent reasoning capabilities [4], as they allow the detection of previously unknown malicious traffic. The solution proposed in this research could be used for building hybrid NIDS that would use both these new anomaly-based strategies as well as the traditional misuse strategies that are usually built by means of rules. In this sense, the enriched middleware designed could be extended to include these new types of reasoning.

The use of the JADE multi-agent MAS platform makes this work useful only for IoT scenarios where the nodes where the agents will run have the adequate processing capabilities to run such relatively heavy software artifacts. While many wireless sensor networks with constrained devices cannot use this solution, many other use cases, as home or industrial scenarios, as well as cloud-based IoT architectures, can benefit from this approach.

An important consideration when building distributed NIDS solutions based on multi-agent systems is the concern about security and trust of the NIDS solution. The existence of different software agents, with their own set of behaviors and mobility capabilities, as well as the set of communications between agents that occur during the MAS NIDS operation poses a number of new threats that, if not adequately addressed, may involve a new source of security issues for the network. In order to build security and trust into multi-agent platforms, the usual approach consists in building encrypted communication channels as well as implementing trust assurance mechanisms for the agents. In this sense, the JADE-S add-on [19] includes functionalities for message encryption and signature, agent actions authorization against agent permissions, and user authentication. This add-on has been successfully used in IoT scenarios [20] but demands even more resources from the nodes where the agents using it live. A more recent approach for building security and trust into multi-agent systems is the use of techniques based on blockchain technologies [21]. These techniques have been recently applied to a multi-agent Network Intrusion Detection System using the JADE MAS platform with promising results [22]. This technology allows to secure the ACL communications among agents while regulating this process by implementing a smart contract mechanism for trust management. Also, blockchain technologies do not imply a great overload for the nodes in an IoT environment and have been applied in scenarios including constrained devices [23]

## 7 Conclusions

The enriched multi-agent middleware presented in this paper eases the incremental development and debugging of rule-based multi-agent systems. The strategy used for integrating the rule engine into the agent made it possible to obtain a more flexible and faster solution than other similar ones, what is an advantage in knowledge-intensive multi-agent applications, as is the case of Intrusion Detection Systems in IoT environments.

There are a number of limitations regarding this work, as stated in the previous section. Further work is needed in order to build the possibility of including agents which use modern machine learning and deep learning algorithms as long as the rule-based ones in order to build hybrid NIDS. Security and trust mechanisms should be built into the multi-agent system, in this sense, the recent application of blockchain technologies to multi-agent-based IDS is a promising research area. Future work also includes the implementation of more complex collaboration and coordination mechanism for the *rule-based agents*, and the use of Semantic Web formalisms (RDF, SWRL, etc.) for representing rules and facts into the agents.

**Funding** This work was supported by Junta de Castilla y León, Spain [grant number LE078G18].

**Publisher's Note** Springer Nature remains neutral with regard to jurisdictional claims in published maps and institutional affiliations.

## Authors and Affiliations

**Francisco José Aguayo-Canela[1] · Héctor Alaiz-Moretón[1] · María Teresa García-Ordás[1] · José Alberto Benítez-Andrades[2]**  **· Carmen Benavides[2] · Isaías García-Rodríguez[1]**

　　Francisco José Aguayo-Canela
　　francisco.aguayo@ieee.org

　　Héctor Alaiz-Moretón
　　hector.moreton@unileon.es

　　María Teresa García-Ordás
　　mgaro@unileon.es

　　Carmen Benavides
　　carmen.benavides@unileon.es

　　Isaías García-Rodríguez
　　isaias.garcia@unileon.es

[1] SECOMUCI Research Group, Escuela de Ingenierías Industrial e Informática, Universidad de León, Campus de Vegazana s/n, C.P. 24071 León, Spain

[2] SALBIS Research Group, Department of Electric, Systems and Automatics Engineering, University of León, Campus of Vegazana s/n, León, 24071 León, Spain